\documentclass[sn-nature,iicol]{sn-jnl}
\usepackage{graphicx}%
\usepackage{multirow}%
\usepackage{amsmath,amssymb,amsfonts}%
\usepackage{amsthm}%
\usepackage{mathrsfs}%
\usepackage[title]{appendix}%
\usepackage[table]{xcolor}%
\usepackage{textcomp}%
\usepackage{manyfoot}%
\usepackage{booktabs}%
\usepackage{algorithm}%
\usepackage{algorithmicx}%
\usepackage{algpseudocode}%
\usepackage{listings}%
\usepackage{placeins}
\usepackage{arydshln}
\usepackage{float}
\usepackage{tabularx}
\usepackage{soul}
\let\OLDthebibliography\thebibliography
\renewcommand\thebibliography[1]{
  \OLDthebibliography{#1}
  \setlength{\parskip}{0pt}
  \setlength{\itemsep}{0pt plus 0.3ex}
}
\begin{document}
\title[Watching Physics: 
the Generative Science of Matter and Motion]
{\sffamily\bfseries
\hspace*{3.0cm}Watching Physics: \\
the Generative Science of Matter and Motion}
\author[1]{\fnm{Hagen}     \sur{Holthusen}}  \email{hagen.holthusen@fau.de}
\author[2]{\fnm{Kevin}     \sur{Linka}}      \email{linka@ame.rwth-aachen.de}
\author[1,3]{\fnm{Ellen}     \sur{Kuhl}}     \email{ekuhl@stanford.edu}

\affil[1]{\small{Institute of Applied Mechanics,
University of Erlangen-Nuremberg, 91058 Erlangen, Germany}}

\affil[2]{\small{Computational Mechanics in Medicine,
RWTH Aachen University, 52074 Aachen, Germany}}

\affil[3]{\small{Department of Mechanical Engineering,
Stanford University, Stanford, CA 94305, United States}}
\abstract{
\textit{Can we learn the physics of matter in motion directly from images and video—and trust it?} Answering this question requires integrating experiments, physics-based simulation, and data across traditionally separate disciplines. Much of this knowledge is visual and temporal rather than textual: images and videos encode structure, dynamics, and causality that equations alone cannot fully capture. Recent generative models produce compelling visual content, yet they rely on observational data and often lack physical validity. Here we show that generative video models gain scientific value when they couple visual data with experiments and high-fidelity simulations. Using deformation mechanics as a testbed, we study three systems of increasing complexity—rubber compression, can crushing, and cardiac motion—and identify regimes in which visual learning succeeds, fails, and requires mechanistic supervision. When physics manifests in visible kinematics, generative models recover measurable quantities such as surface strain; when internal state variables dominate, visual plausibility no longer ensures physical admissibility. We propose that this convergence defines a new frontier, the \textit{Generative Sciences of Matter and Motion}, which unifies Simulogenics, Physiogenics, and Materiogenics. These physics-grounded foundation models can turn visual generation into a scientific instrument for inference, prediction, and design of matter in motion.}
\maketitle
\section*{\sffamily{\bfseries{Motivation}}}
\noindent
For centuries, 
science has sought to \textit{explain the world} 
through equations of matter and motion. 
From Hooke’s law of elasticity~\cite{hooke1678} 
to Newton’s laws of dynamics~\cite{newton1687} 
and the Navier–Stokes equations of fluids~\cite{navier1822,stokes1845}, 
physics has advanced by abstracting nature into mathematical form. 
These abstractions have enabled predictive simulation 
and transformed computers into laboratories for virtual experimentation~\cite{vonneumann1945}. 
In parallel, 
artificial intelligence has evolved 
from symbolic reasoning~\cite{turing1950,mccarthy1955} 
to deep neural networks that generate text, images, and video~\cite{lecun2015,goodfellow2014}. 
Information theory has formalized this transition 
from description to representation~\cite{shannon1948}, 
and modern generative models 
now learn the latent structure of the data themselves~\cite{ho2020,brown2020}. 
Together, these developments mark a shift 
\textit{from explaining} nature \textit{to generating} it
and unify physics and AI into the {\it{Generative Sciences of Matter and Motion}}. 
\begin{figure*}[h]
    \centering
    \includegraphics[width=1.0\textwidth]{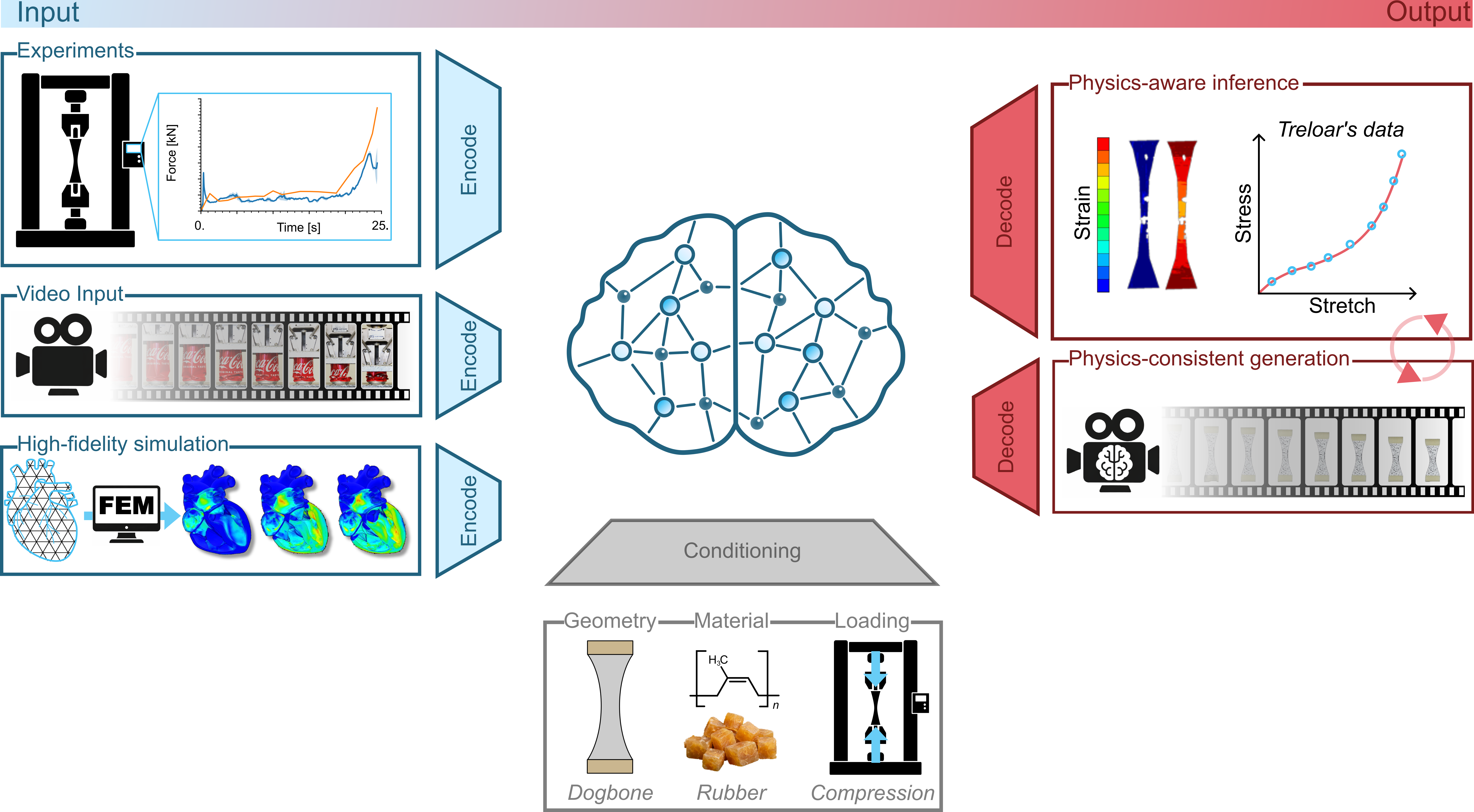}
\caption{{\sffamily{\bfseries{Generative Science of Matter and Motion.}}}
Conceptual workflow of the proposed generative-video-AI framework. On the input side, the model is trained on diverse video streams and learns deformation patterns of materials and structures under external loading from visual dynamics. To improve physical fidelity and quantitative consistency, this learning process is further reinforced with experimentally observed responses and video sequences from high-fidelity computer simulations. The resulting latent representation enables a material and structural understanding and supports inference of physically plausible generated simulations. On the output side, the framework supports the generation and inference of physically plausible motion and deformation across scales. This includes not only visually realistic video sequences, but also quantitatively meaningful predictions that can be evaluated, conditioned, or refined through physics-aware inference. The resulting paradigm bridges observation, simulation, and generation, and enables applications such as inverse material characterization, virtual experimentation, and predictive modeling of complex systems.}   
\label{fig:Overview}
\end{figure*}\\[6.pt]
\noindent
{\sffamily{\bfseries{Simulogenics--From equations to generation.}}}
Simulogenics establishes the generative science of physical simulation, 
where models learn to produce 
physically consistent trajectories of matter and motion. 
Classical mechanics has grounded this effort 
in variational principles and conservation laws~\cite{hestenes1973,newton1687}, 
and continuum theories have formalized 
the dynamics of complex media~\cite{navier1822,stokes1845}. 
The finite element method has translated these equations 
into computational frameworks for predictive simulation~\cite{zienkiewicz1977}, 
and digital twins have extended this paradigm 
toward real-time predictive systems~\cite{grieves2014}. 
Modern generative models now invert this process 
and learn to generate solutions directly from data~\cite{sohl2015,ho2020}. 
This convergence suggests 
that simulation itself can be learned 
and bridge first-principles physics 
with data-driven generation
to enable \textit{learning simulation without equations}.\\[6.pt]
\noindent
{\sffamily{\bfseries{Physiogenics--From observation to governing laws.}}}
Physiogenics defines the generative science 
of inferring physical laws from visual observations of matter in motion. 
Foundational theories of perception 
have framed vision as an inverse problem 
that reconstructs the world from images~\cite{marr1982,gibson1979}, 
and statistical physics has demonstrated 
that complex dynamics follow structured stochastic laws~\cite{kolmogorov1941}. 
Quantum mechanics has further recast physical behavior 
in probabilistic terms~\cite{feynman1948}. 
Recent advances in generative video models 
now encode rich dynamical priors 
directly from visual data~\cite{ho2022video,ha2018world}. 
This perspective suggests 
that physical laws need not be prescribed explicitly, 
but can instead emerge from observation
which elevates video to a primary modality for scientific discovery, 
and enables \textit{learning physics from vision}. \\[6.pt]
\noindent
{\sffamily{\bfseries{Materiogenics--From structure to functional design.}}}
Materiogenics introduces the generative science 
of designing matter with prescribed properties and functions. 
Classical materials science has established 
structure–property relationships 
as the foundation for materials selection 
and engineering design~\cite{ashby1996}. 
Advances in topology optimization 
have enabled systematic inverse design 
of high-performance structures~\cite{bendsoe2003}. 
At the same time, 
foundational work in biophysics has connected physical laws 
to the organization of living matter 
and highlighted the interplay 
between physics, biology, and function~\cite{schrodinger1944}.
Recent advances in self-supervised representation learning demonstrate how we can capture structured physical dynamics from data, 
which results in predictive world models grounded in temporal coherence and interaction~\cite{assran2025vjepa2}.
Generative models now navigate high-dimensional design spaces 
and propose novel material configurations~\cite{goodfellow2014,brown2020}. 
Yet, these models increasingly benefit from the encoding of simple physical simulations, such as rigid-body motions, which provide low-dimensional structure to ensure physically consistent generation~\cite{liu2024physgen}.
This shift reframes materials design as a generative process and enables \textit{learning matter by design}. \\[6.pt]
\noindent
In the context of mechanics and materials science, these advances prompt a fundamental question: 
\textit{Can models trained on unstructured video observations alone infer the underlying physics that govern matter and motion?}
Although large-scale datasets contain extensive real-world dynamics, they rarely encode constitutive response, state variables, loading paths, or boundary conditions explicitly. Computational simulations provide these quantities with precision; yet, current generative models rarely incorporate them. As a result, existing approaches achieve strong visual realism but often lack quantitative reliability in physically demanding scenarios. \\[6.pt]
\noindent
{\sffamily{\bfseries{Outline.}}}
This perspective operationalizes \textit{Physiogenics} by using deformation mechanics as a testbed to determine whether generative video AI can infer physics beyond visual realism. We establish that Physiogenics provides the missing link between data-driven observation and generative simulation; it directly connects to \textit{Simulogenics}, and opens the path toward \textit{Materiogenics}. The framework integrates visual observations, experiments, and high-fidelity simulations to learn a latent representation that encodes the governing mechanics of matter and motion and supports generation and quantitative inference (Figure~\ref{fig:Overview}). 
We instantiate this paradigm through a staged benchmark with three scenarios of increasing complexity: 
(i) uniaxial deformation, where the model recovers kinematics in the spirit of digital image correlation and compares them to finite element references; 
(ii) nonlinear thin-walled collapse in can crushing, validated against experiments and simulations; and 
(iii) cardiac deformation over one heartbeat as a high-dimensional biomechanical test case. 
These examples map input to latent representation to output, quantify capabilities, expose limitations, and define the requirements for 
\textit{learning physics from vision}, 
\textit{learning simulation without equations}, and 
\textit{learning matter by design}.
\section*{\sffamily{\bfseries{Integrating experiment, generation, and simulation}}} 
In the following three exploratory studies, we consistently employed the generative video artificial intelligence model \textit{Sora} by OpenAI to generate the visual data.
\\[10.pt]
{\sffamily{\bfseries{Exploratory\,Study\,I: Compressing a rubber block.}}}
In the first study, we explore the deformation and strain fields of a rubber block under uniaxial compression. Rubber compression distributes the mechanical loading smoothly across the entire body. Large deformations, near-incompressibility, and geometric nonlinearity dominate the response; yet, the overall deformation remains smooth, continuous, and surface-visible. 

\begin{figure*}[h]
    \centering
    \includegraphics[width=1.0\textwidth]{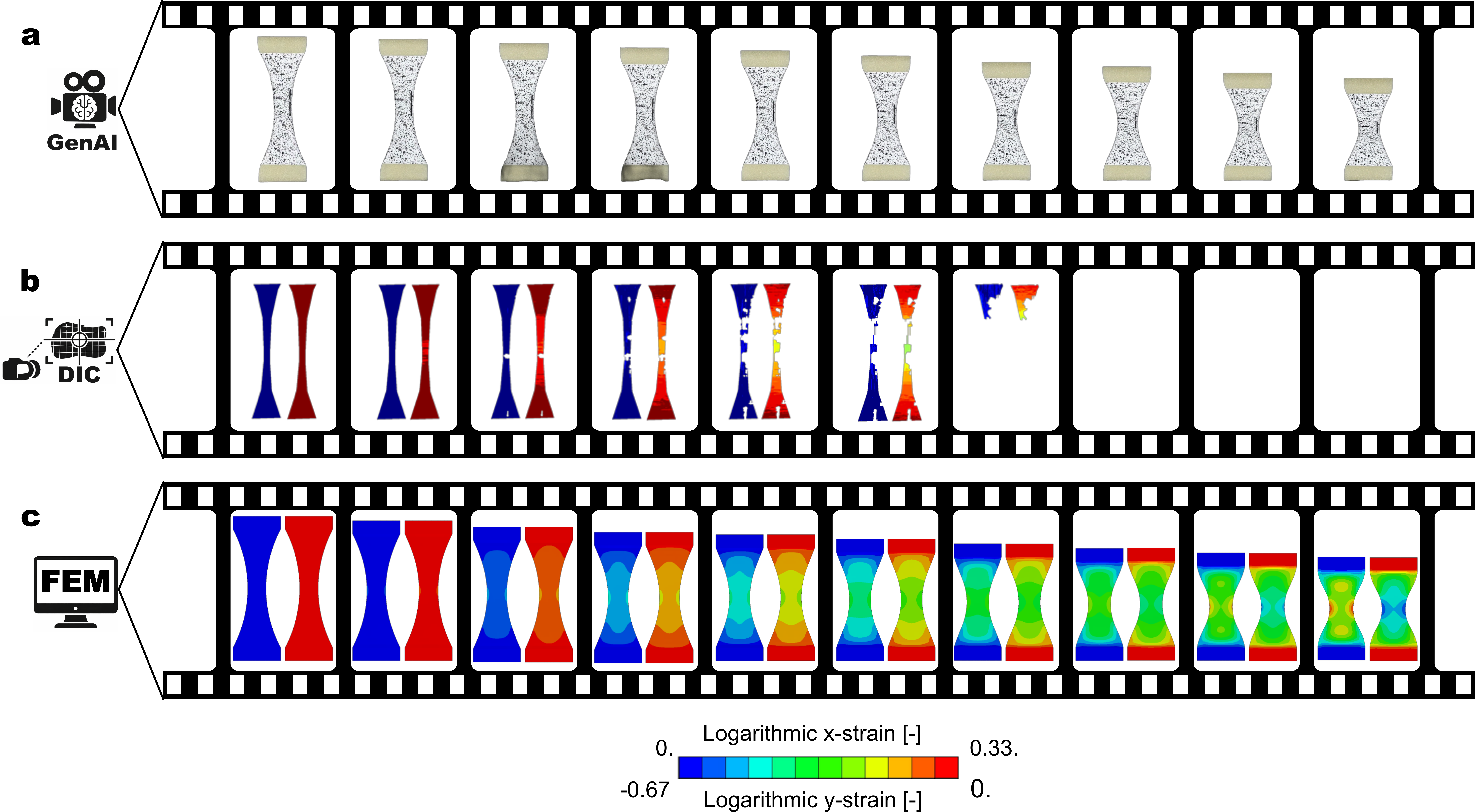} 
\caption{{\sffamily{\bfseries{Exploratory Study I: Compressing a rubber block.}}}
Comparison of 
generative AI video,
digital image correlation, and 
finite element simulation 
for the displacement-controlled compression of a rubber block. 
The generative AI video sequence captures the macroscopic hourglass-shaped deformation under axial compression (top). 
From the video, we extract logarithmic strain fields,
including axial strain (left) and transverse strain (right), 
from digital image correlation (middle).
The finite element simulation (bottom)
based on a third-order Ogden hyperelastic model 
predicts compressive axial logarithmic strains (left) and tensile transverse logarithmic strains (right). 
During the early stages of deformation, 
digital-image-correlation-derived strains from the generated video agree well with the finite element simulation
and surface-visible kinematic patterns are captured with quantitative fidelity.
During larger stages, 
feature tracking deteriorates due to limited temporal coherence in the generated sequence. 
Together, this comparison highlights the complementary roles of 
experiment for measurable surface kinematics, 
simulation for full-field mechanical consistency, and 
generation for scalable, variability-rich virtual experiments.}  
\label{fig:Rubber}
\end{figure*}

Figure~\ref{fig:Rubber} 
compares three complementary perspectives 
of rubber block compression:
a generative AI video (top), 
digital image correlation applied to images of the video (middle), and 
a finite element simulation based on a third-order Ogden hyperelastic model (bottom), with details in Table~\ref{tab:RubberCompress}.
Figure~\ref{fig:RubberDimension} provides additional details about the region of interest for digital image correlation and the dimensions for the finite element simulation.

The \textit{generative AI video} establishes the visual baseline: 
The specimen contracts axially, expands laterally, and forms the characteristic hourglass shape of an incompressible hyperelastic solid. 
Because deformation follows smooth kinematics 
rather than localized failure, 
visible geometry encodes the essential mechanics 
and supports quantitative post-processing.

The \textit{digital image correlation} 
with \texttt{PYVALE}~\cite{hirst2026pyvale} allows us to 
extract logarithmic strain fields directly from the synthetic video. 
The strain fields display the expected trends with
compressive axial strain in blue and tensile transverse strain in red,
with a heterogeneous strains distribution along the length of the specimen. 

The \textit{finite element simulation} provides the mechanical reference. We model the material as incompressible hyperelastic, fit its parameters \cite{steinmann12} to classical rubber data~\cite{treloar44}, and discretize it with reduced-integration hexahedral elements. The simulation resolves the full deformation gradient, strain fields, and stress response at every integration point. Unlike the experiment or the generated video, the finite element simulation provides direct access to internal fields: principal stretches, strain energy density, and reaction forces emerge as consistent solutions of the balance equations. Yet, these predictions depend on modeling choices: the Ogden form, enforced incompressibility, and idealized boundary conditions. While the finite element simulation offers completeness and interpretability, it remains a calibrated hypothesis.

In the side-by-side comparison, 
in early compression, axial and transverse strains 
from the digital image correlation agree closely with the finite element simulation: deformation symmetry, compression magnitude, and lateral expansion match the expected hyperelastic response. This agreement identifies a regime in which the generative video recovers surface kinematics at a level useful for calibration. In this regime, generative models can function as virtual experiments—generate ensembles, provide strain fields via digital image correlation, and calibrate constitutive models without repeated laboratory testing. Their stochastic variability further mirrors experimental noise and enables controlled data augmentation. Limitations appear as deformation increases. The digital image correlation deteriorates in the generated sequence: texture evolution disrupts feature correspondence, strain recovery fails, and temporal coherence at the material point level breaks down. While the video model captures the overall shapes, it fails to accurately reproduce motion in the form of particle trajectories.
\begin{figure*}[h]
    \centering
    \includegraphics[width=1.0\textwidth]{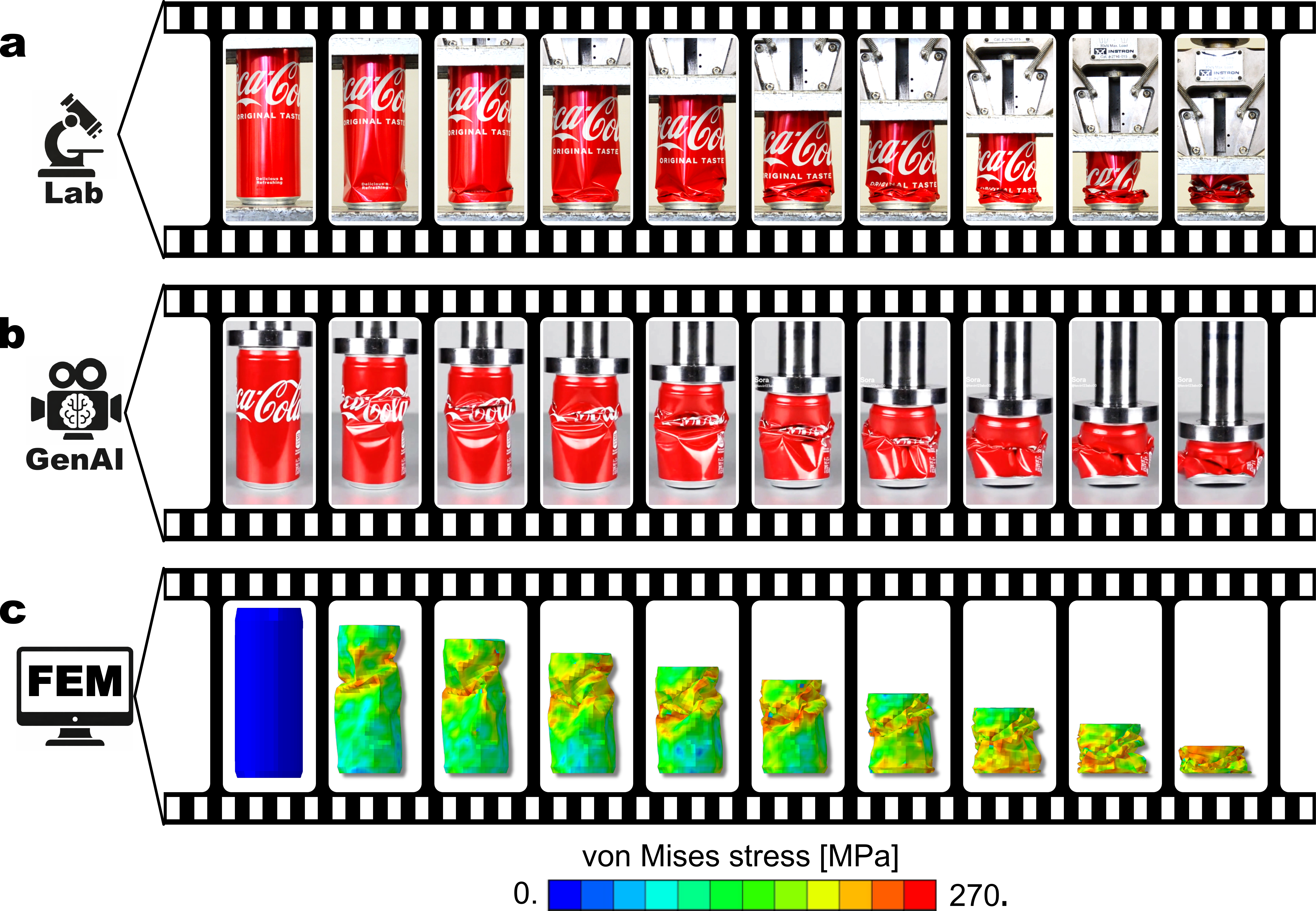} 
\caption{{\sffamily{\bfseries{Exploratory Study II:  Crushing a can.}}}
Comparison of 
laboratory experiment,
generative AI video, and 
finite element simulation 
for the axial compression of a thin-walled AA-3104-H19 aluminum can.
The laboratory experiment provides the exact collapse sequence and measured reaction forces in real life (top). 
The generative AI video sequence captures the global kinematic evolution of progressive buckling and folding (middle).
The finite element simulation 
resolves the von Mises equivalent stress and 
captures plastic hinge formation and stress localization.
The laboratory experiment offers physical authenticity, 
but limited access to internal fields; 
the generative model reproduces visually plausible dynamics and illustrates the potential for scalable, language-driven access to structural scenarios;
the simulation provides full-field mechanical insight under explicit modeling assumptions. 
Together, this comparison delineates the complementary strengths of observation, computation, and generation in the study of structural instability.}  
\label{fig:ColaCan}
\end{figure*}

Taken together, the comparison delineates the individual roles of the three modalities: 
the generated videos capture accessible kinematics;
the digital image correlation extracts measurable surface fields; and 
the finite element simulation provides full mechanical consistency. When physics manifests in visible kinematics, the generative video approaches quantitative usefulness; yet, without additional physical constraints, it does not ensure consistent material mapping or constitutive admissibility.

Within the \textit{Generative Sciences of Matter and Motion}, this example highlights a promising intermediate regime: When physics is strongly expressed in visible kinematics, generative video can already approximate experimentally relevant data streams. By coupling such models with simulation-informed supervision and enforcing temporal and physical consistency, virtual experiments can become a practical bridge between laboratory observation and mechanistic simulation and reduce the experimental burden while preserving mechanical interpretability. \\[10.pt]
{\sffamily{\bfseries{Exploratory Study II: Crushing a can.}}}
In the second study, we explore the crushing of a can under compression. Thin-walled structures concentrate mechanics into geometry. Their load-bearing capacity emerges from the interplay of elastic stiffness, plastic hardening, instability, and contact. The axial crushing of an aluminum beverage can is therefore a compact yet demanding benchmark: simple in appearance; yet, governed by nonlinear shell kinematics and progressive buckling.
This makes the can an instructive counterpoint to the hyperelastic rubber block: here, much of the relevant physics is not directly observable.

Figure~\ref{fig:ColaCan} compares three complementary views of can crushing: 
a laboratory experiment (top), 
a generative AI video (middle), and 
a finite element simulation (bottom),
with details in Table \ref{tab:ColaCan}.
Figure~\ref{fig:ColaCanDimension} defines the shell geometry and numerical setup, and Figure~\ref{fig:ColaCanCurves} reports the corresponding force–time response. The sequence captures progressive buckling and folding of a thin-walled aluminum can, a canonical instability problem governed by shell kinematics, plasticity, and contact.

The \textit{laboratory experiment} establishes the physical baseline: 
It captures real geometry, material imperfections, and manufacturing variability of the AA-3104-H19 alloy. The force–time curve exhibits the characteristic signature of collapse, an initial elastic regime, a peak load, followed by abrupt drops and oscillations associated with fold initiation and progressive plastic hinging. The experiment provides authenticity and irreducible complexity, but limits access to internal fields. We observe deformation and reaction forces, but not stress, plastic strain, or energy redistribution.

The \textit{generative AI video} reproduces the global kinematic narrative of collapse. The can shortens, buckles, and forms folds in a visually plausible sequence. This representation captures the emerging morphology of the instability, but does not expose internal variables or enforce equilibrium and constitutive constraints. The evolution remains appearance-consistent rather than physics-consistent: the model interpolates learned deformation patterns, but it does not guarantee correct responses under modified loading, material parameters, or boundary conditions.

The \textit{finite element simulation} provides the mechanical reference. We construct the shell geometry, prescribe an isotropic elastoplastic constitutive law, and solve the problem with an explicit time integration using Abaqus/Explicit. The simulation resolves stress fields, plastic strain localization, fold formation, and reaction forces at each time increment. It exposes the internal state—momentum balance, constitutive response, and contact interactions—throughout collapse. Yet, the predictions depend on modeling choices: idealized geometry, assumed material law, friction parameters, and finite resolution. The FEM delivers interpretability and control, but remains a calibrated hypothesis.

Taken together, the comparison delineates the roles of the three modalities: the experiment captures physical reality with full complexity but limited observability;
the generative model captures accessible kinematics and enables scalable exploration of collapse scenarios; and
the finite element simulation resolves full mechanical consistency under explicit assumptions. 
\begin{figure*}[h]
    \centering
    \includegraphics[width=1.0\textwidth]{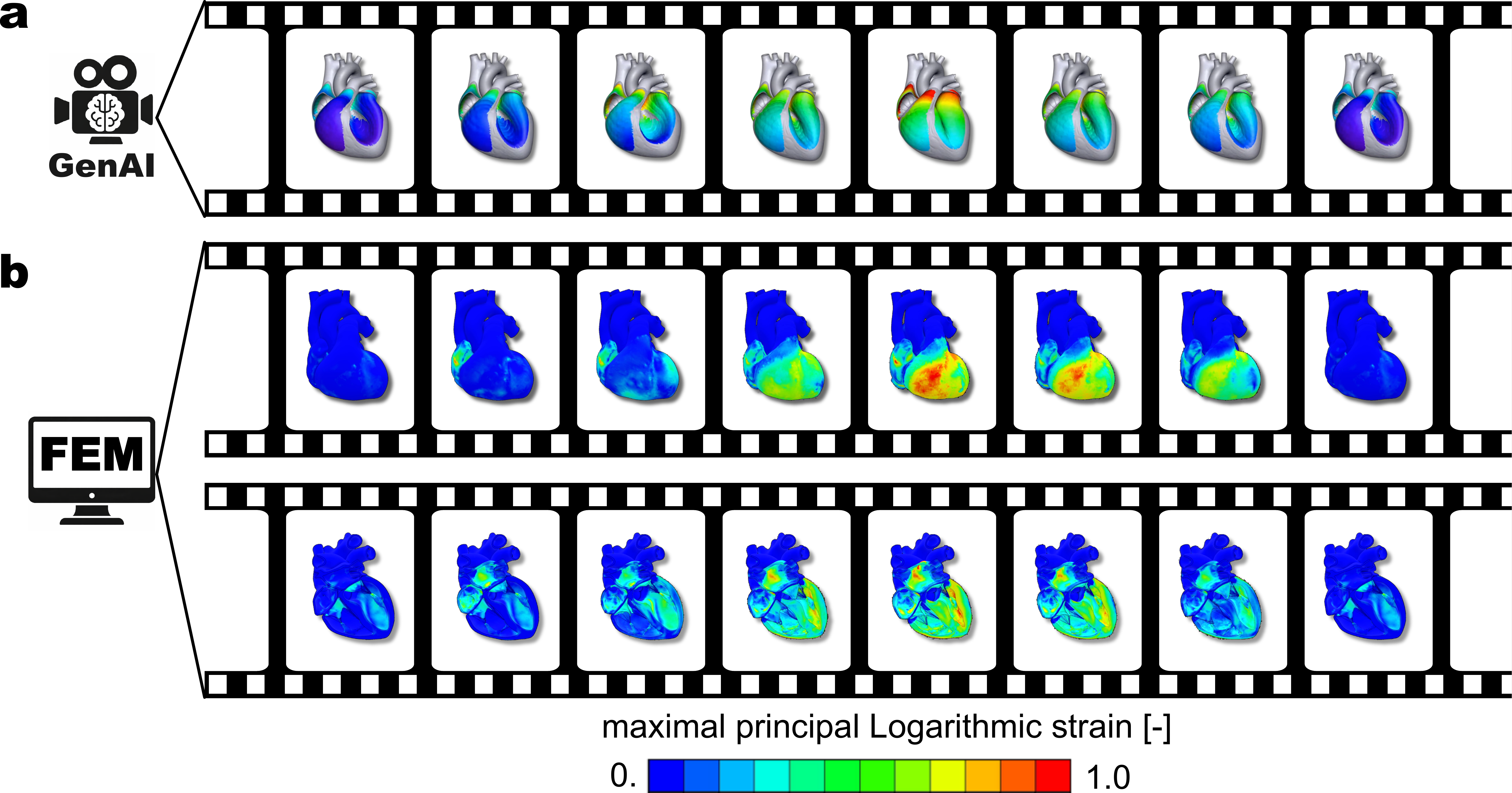} 
\caption{{\sffamily{\bfseries{Exploratory study III: 
Beating of the human heart.}}}
Comparison of generative AI video and
finite element simulation throughout one cardiac cycle. 
The generative AI video sequence reproduces global ventricular contraction and relaxation (top).
The finite element simulation 
highlights the maximum principal logarithmic strain profile 
in front view (middle) and four-chamber view (bottom).
The generative model does not accurately capture the geometry of a healthy human heart. It qualitatively reproduces the global kinematics of cardiac contraction, but does not enforce mechanical admissibility; the displayed contour plots are illustrative rather than physically derived. 
The finite element model resolves cardiac motion 
from the governing equations of the momentum balance 
with detailed constitutive descriptions of myocardial tissue
and provides full-field access to the deformation. 
Together, this comparison highlights the complementary strengths of 
generative modeling, scalability, multi-modal data integration, and potential for patient-specific adaptation, and
physics-based simulation, mechanistic rigor and interpretability, and  underscores the importance of realistic clinical images and 
physics-grounded generative architectures 
for reliable medical predictions.}
\label{fig:Heart}
\end{figure*}
Unlike the rubber case, the governing physics do not fully manifest in visible kinematics; instead, instability, plasticity, and contact require internal state variables. As a result, the generative video reproduces the appearance of collapse, but does not ensure mechanical admissibility. 

Within the \textit{Generative Sciences of Matter and Motion}, this example defines a regime in which visual learning alone falls short and highlights the need for physics-grounded generative models that integrate observation, simulation, and generation into a unified framework.  \\[10.pt]
{\sffamily{\bfseries{Exploratory study III: Beating of the human heart.}}}
While rubber compression probes smooth hyperelasticity, 
and can crushing probes structural instability, 
the beating human heart probes the frontier of computational mechanics. 
Cardiac function couples nonlinear anisotropic material behavior, active contraction, complex boundary conditions, and multi-scale structural organization within a moving, perfused organ. Unlike the previous examples, this study is deliberately forward-looking. It illustrates not what generative models already achieve reliably, but what becomes conceivable when physics-based simulation and foundation models converge.

Figure~\ref{fig:Heart} compares two complementary views of cardiac deformation over one heartbeat: generative AI video (top) and high-fidelity finite element simulation in front view (middle) and four-chamber view (bottom). The sequence captures ventricular contraction and relaxation, while the simulation resolves the corresponding maximal principal logarithmic strain fields.

The \textit{generative AI video} establishes the kinematic narrative. The 
heart fills, the ventricles contract, and the heart relaxes in a visually coherent cycle. This representation captures the global dynamics of cardiac motion and reflects patterns learned from large-scale visual data. 
However, at a closer look, the big vessels of the generated heart geometry do not accurately reproduce the geometry of a healthy human heart. The simulation
does not expose internal variables or enforce mechanical admissibility. The contour fields remain illustrative, with unphysiological high-strain region at the relatively rigid base, and the evolution reflects appearance-consistent motion rather than a solution of governing equations.

The \textit{finite element simulation} provides the mechanistic reference. It resolves one cardiac cycle from the equations of motion using anatomically accurate geometry, anisotropic constitutive laws, and active contraction \cite{baillargeon14}. The simulation computes deformation from first principles and exposes full-field quantities--stresses, strains, energy densities, and reaction forces--throughout the myocardium \cite{peirlinck21}. This approach delivers quantitative consistency and interpretability, but it requires detailed imaging, parameter calibration, and substantial computational effort. Model construction, personalization, and repeated calibration limit scalability.

The comparison highlights a fundamental gap: Unlike rubber deformation, cardiac mechanics does not manifest fully in visible kinematics. Active stress generation, fiber architecture, and internal fluid flow govern the response and require internal state variables. As a result, the generative model reproduces the motion of the heartbeat, but not its constitutive structure.
At the same time, this example defines a clear opportunity: High-fidelity simulations can provide structured supervision--dense fields of strain, stress, and activation--that guide generative models toward physically admissible representations. In such a hybrid setting, physics defines the admissible space of cardiac dynamics, while generative models provide scalability and multimodal integration. This combination enables a new paradigm for patient-specific modeling where sparse clinical data can condition a physics-informed generative model that reconstructs a high-resolution, mechanically consistent digital twin.

Within the \textit{Generative Sciences of Matter and Motion}, this case marks the frontier. The generative model captures global dynamics but lacks mechanistic grounding; the simulation provides mechanistic rigor but limits accessibility. Bridging both defines the path forward toward scalable, physics-consistent modeling of living systems. 
\section*{\sffamily{\bfseries{From visual realism to physical validity}}}
Visual realism alone does not guarantee physical truth; models must capture mechanism to achieve validity.\\[6.pt]
{\sffamily{\bfseries{From appearance to mechanism}}}
Generative modeling must move from appearance to mechanism to become scientifically meaningful. 
Models that learn only from visual data reproduce motion, but models that integrate experiments, simulations, and mechanical structure capture its governing principles. 
The three case studies define this transition across increasing physical complexity. 
In rubber compression (Figure~\ref{fig:Rubber}), visible kinematics encode enough information to recover surface strain fields and support calibration in an intermediate regime. 
In can crushing (Figure~\ref{fig:ColaCan}), visual information no longer suffices, because instability, plasticity, and contact depend on internal state variables. 
In cardiac motion (Figure~\ref{fig:Heart}), physiological relevance requires mechanistic consistency from the outset. 
Across these regimes, the decisive question is not whether generated motion \textit{looks} realistic, but whether it remains interpretable in terms of deformation, loading, and constitutive response.\\[6.pt]
{\sffamily{\bfseries{From plausibility to validity}}}
Visual plausibility must give way to physical validity. 
The crushed can example (Figure~\ref{fig:ColaCan}) exposes this gap most clearly: the generated sequence reproduces the morphology of collapse, yet it does not guarantee mechanical admissibility. 
This discrepancy shows why perception-based evaluation falls short. 
Validation must rely on physics-based criteria, including admissible deformation paths, consistency with governing principles, and agreement with experimental or simulated responses. 
These criteria determine whether a model captures physical behavior rather than reproducing its appearance. \\[6.pt]
{\sffamily{\bfseries{From prediction to design}}}
Generative models can move from prediction to design. 
The cardiac example (Figure~\ref{fig:Heart}) illustrates this opportunity: patient-specific data can condition a generative model and guide the exploration of individualized biomechanical responses. 
By inverting learned relationships between structure, motion, and mechanics, we can generate configurations that realize targeted physical behavior. 
This perspective extends classical inverse design by combining multimodal inputs with rapid exploration in a physics-consistent design space.\\[6.pt]
{\sffamily{\bfseries{From access to understanding}}}
Generative models can move from access to understanding in physical reasoning. 
The rubber, can, and heart studies span a progression from controlled benchmarks to complex living systems, while exposing the limits of visual learning. 
Highly nonlinear phenomena such as structural instability remain difficult to characterize reliably from appearance alone, as the can example shows. 
More broadly, current models depend on curated datasets and often fail in extrapolation. 
At the same time, physics-grounded supervision can lower the barrier to physically meaningful surrogates and broaden access to simulation-based reasoning. 
Such models can enable more users to explore, interpret, and query complex mechanical systems without specialized numerical workflows. 
Realizing this vision will require robust, physics-informed foundation models that operate across domains and scales and support both scientific discovery and engineering design.\\[6.pt]
\noindent
Taken together, these advances mark the transition from visual realism to physical validity.
\section*{\sffamily{\bfseries{Conclusion and Outlook}}}
When AI not only sees the world but obeys its laws, physics enters the generative age. We show that generative models gain scientific value when they integrate experiments, simulations, and mechanical structure, moving from visual realism to physical validity across regimes of increasing complexity. This shift unifies data and equations into a single framework and turns generative models into tools for simulation, inference, and design. Physics-consistent generative AI can transform how we understand, predict, and engineer matter in motion. \\[6.pt]
\backmatter
\bmhead*{\sffamily{\bfseries{Acknowledgments}}}
We thank Gunnar Possart for his assistance during the can crushing experiments
and Jiang Yao for the heart simulations within the Dassault Systemes Living Heart Project. 
This work was supported by the German Research Foundation DFG 
through CRC/TRR 280 (project ID: 417002380) to Hagen Holthusen and 
through the Emmy Noether Grant (project ID: 533187597) to Kevin Linka, and 
by the National Science Foundation NSF (project CMMI 2320933) and
by the European Research Council ERC (project DISCOVER 101141626) to Ellen Kuhl.
\bmhead*{\sffamily{\bfseries{Conflicts of interest}}} 
All authors declare no financial competing interests.
\bmhead*{\sffamily{\bfseries{Statement of AI-assisted tools usage}}} 
The authors acknowledge the use of \textit{Sora} by OpenAI
for video generation. 
\bmhead*{\sffamily{\bfseries{Availability of data and materials}}} 
All data and materials will be made available upon request. 
\bmhead*{\sffamily{\bfseries{Author contributions}}} 
HH, KL, and EK designed the layout, generated the examples, created the images, and wrote the paper. 

\section*{\sffamily{\bfseries{Appendix}}}
\begin{figure}[h]
    \centering
   \includegraphics[width = 0.3\textwidth]{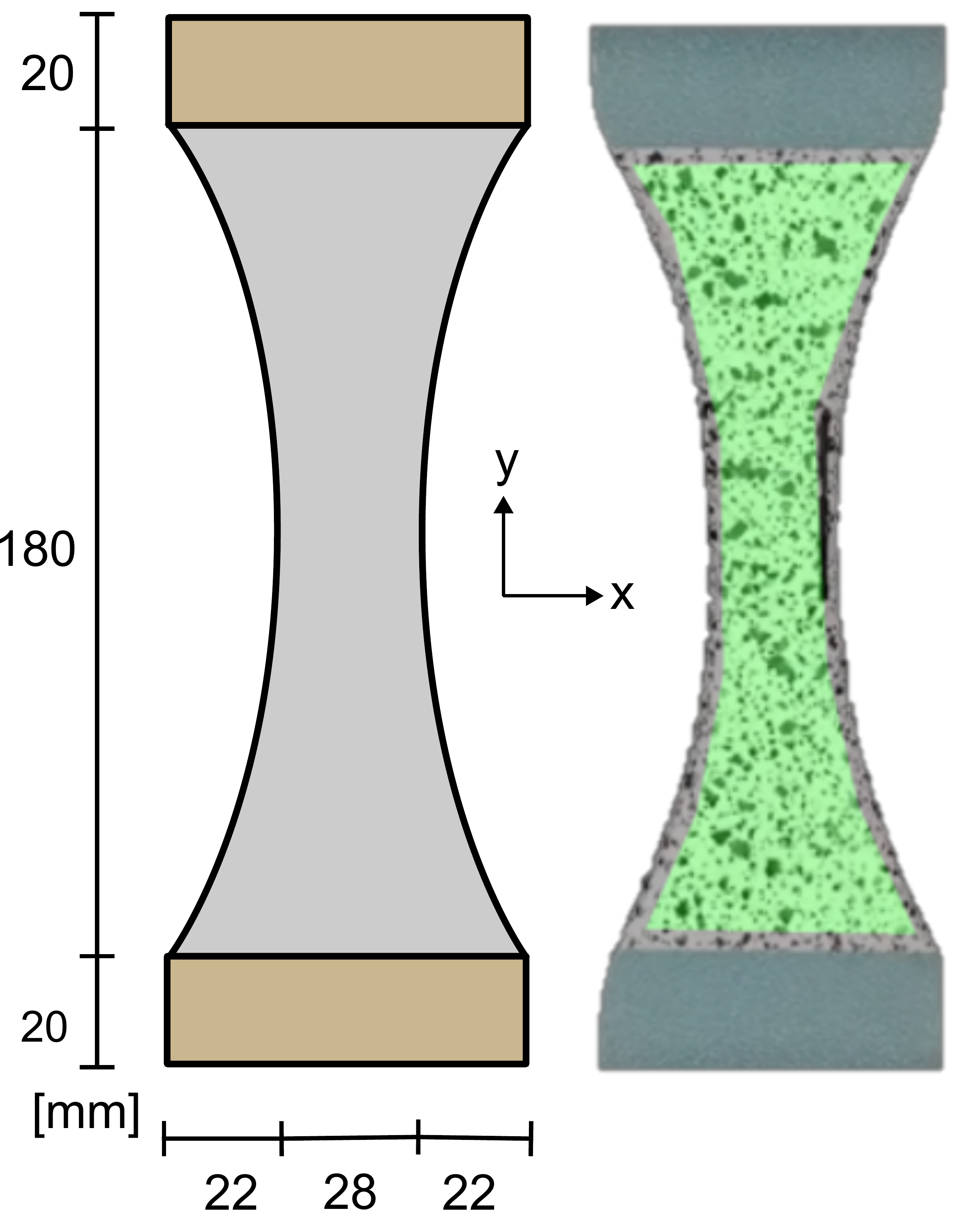}
    \caption{{\sffamily{\bfseries{Exploratory Study I: Virtual compression of a rubber block.}}}
Sample geometry for the finite element simulation (left) and
region of interest for the digital image correlation (right).}    
\label{fig:RubberDimension}
\end{figure}

\begin{figure}[h]
    \centering
   \includegraphics[width = 0.20\textwidth]{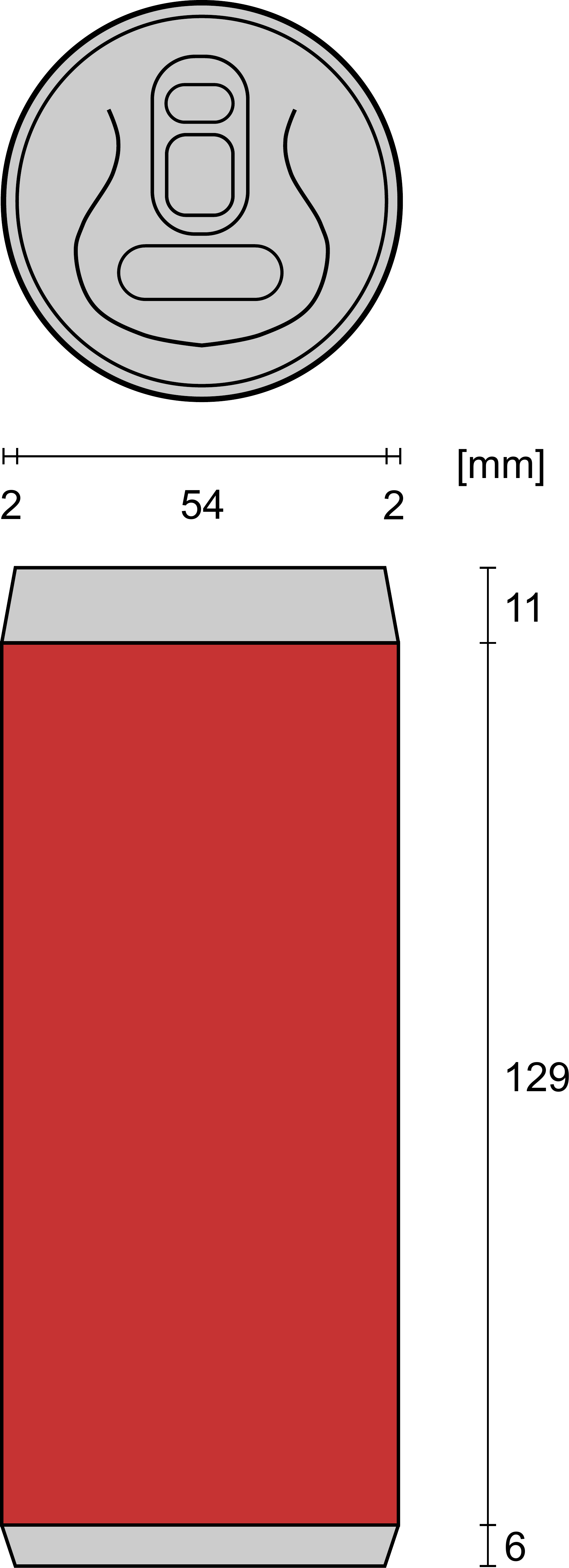}
    \caption{{\sffamily{\bfseries{Exploratory Study II: Virtual crushing of a can.}}}
Sample geometry and numerical for the finite element model simulation.}
\label{fig:ColaCanDimension}
\end{figure}

\begin{figure}[h]
    \centering
    \includegraphics[
        width=0.48\textwidth,
        trim=0 3cm 0 0,
        clip
    ]{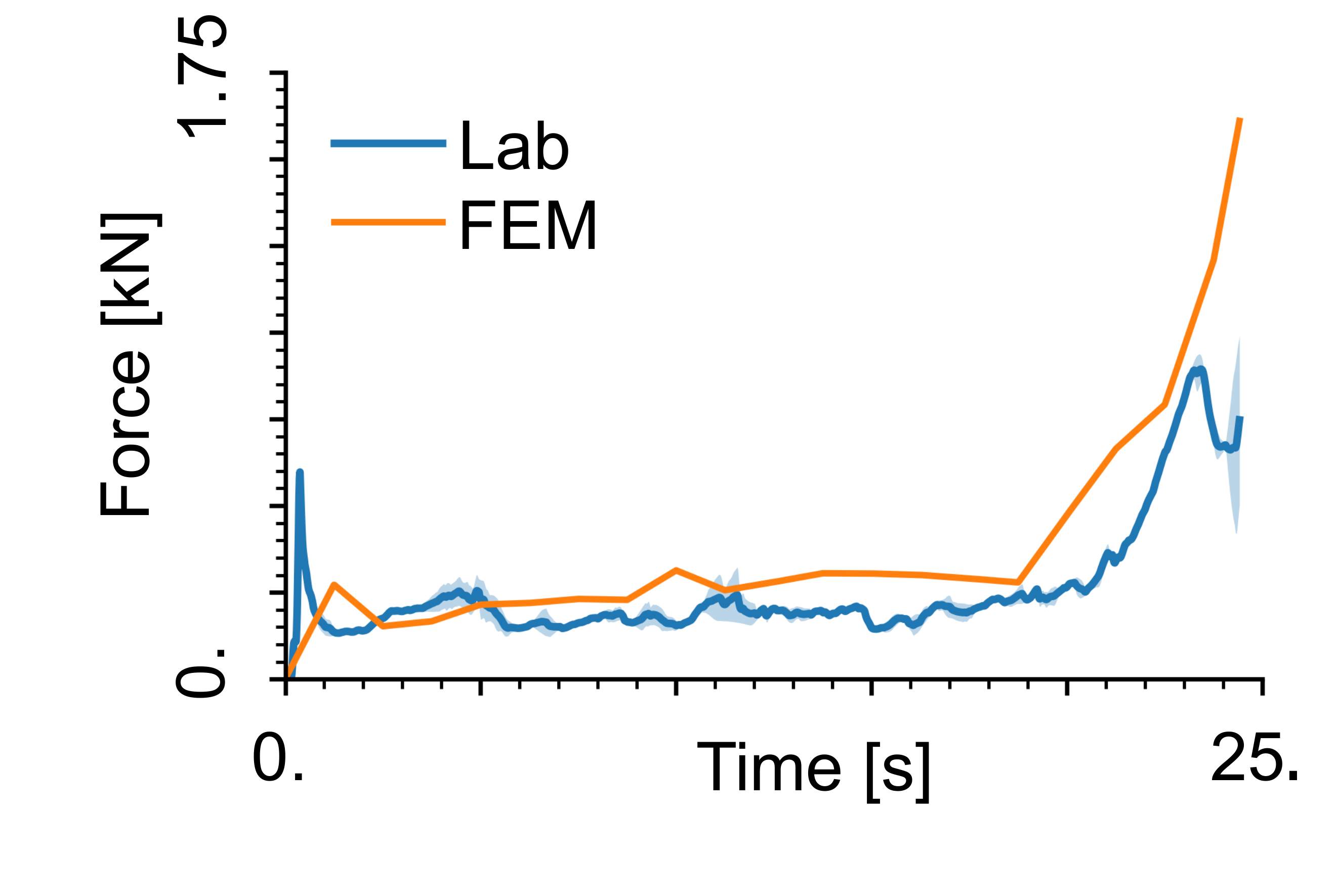}
    \caption{{\sffamily{\bfseries{Exploratory Study II: Structural collapse of crushed can.}}}
The reaction force–time curves from both experiment (blue) and simulation (orange) exhibit the characteristic instability-driven force drops and oscillations during progressive buckling and fold formation.}    
\label{fig:ColaCanCurves}
\end{figure}

Figure~\ref{fig:RubberDimension}
provides additional details 
about the region of interest for digital image correlation 
and the dimensions for the finite element simulation 
of the compression of a rubber block and 
Table~\ref{tab:RubberCompress}
summarizes details about the finite element simulation.
Figure~\ref{fig:ColaCanDimension} defines the shell geometry and numerical setup for the finite element simulation 
of the crushing of a can,
Figure~\ref{fig:ColaCanCurves} reports the experimental and computational force–time response, and 
Table \ref{tab:ColaCan}
summarizes details about the finite element simulation.
\begin{table*}[t]
\centering
\caption{{\sffamily{\bfseries{Exploratory Study I: Material model, loading and numerical parameters for the compression of a rubber block.}}}
The material is described by a third-order Ogden hyperelastic model, with parameters fitted to Treloar’s data \cite{treloar44} following the methodology of \cite{steinmann12}.}
\label{tab:RubberCompress}
\renewcommand{\arraystretch}{1.15}
\small
\begin{tabularx}{\textwidth}{l X}
\toprule
\textbf{Block} & \textbf{Specification} \\
\midrule
\multicolumn{2}{l}{\textbf{Material model}}\\
Hyperelasticity & Ogden model, $N=3$, $D_i=0$ [MPa$^{-1}$] (incompressible) \\
Ogden parameters &
\begin{tabular}{@{}c|ccc@{}}
$\mu_i$ [MPa] & 0.0662 & $5.875\times10^{-12}$ & 0.6249 \\
$\alpha_i$ [--] & 2.875 & 14.221 & 1.0 \\
\end{tabular}
\\
\midrule
\multicolumn{2}{l}{\textbf{Loading program}}\\
Loading type & Displacement-controlled longitudinal compression \\
Total displacement & $u = -70\,\mathrm{mm}$\\
Step duration & $T = 1.0\ \mathrm{s}$ \\
\midrule
\multicolumn{2}{l}{\textbf{Numerical parameters}}\\
Solver / step & Abaqus/Standard; \texttt{*Static, Direct} \\
Number of nodes / elements & 2760 / 1890\\
Element type & \texttt{*C3D8RH} solid element with reduced integration and hourglass control\\
Time incrementation & $\Delta t_0 = 0.1\ \mathrm{s}$, $T = 1.0\ \mathrm{s}$ \\
\bottomrule
\end{tabularx}
\end{table*}
\begin{table*}[t]
\centering
\caption{{\sffamily{\bfseries{Material model, loading and numerical parameters for the crushing of an AA-3104-H19 aluminum can.}}}
The aluminum alloy is described by an isotropic elastoplastic constitutive model with stress–plastic strain data based on \cite{saleh15}.}
\label{tab:ColaCan}
\renewcommand{\arraystretch}{1.15}
\small
\begin{tabularx}{\textwidth}{l X}
\toprule
\textbf{Block} & \textbf{Specification} \\
\midrule
\multicolumn{2}{l}{\textbf{Material model}}\\
Density & $\rho = 2.7\times10^{-9}\ \mathrm{t/mm^3}$ \\
Elasticity & $E = 70\,000\ \mathrm{MPa},\ \nu = 0.33$ \\
Plasticity (isotropic hardening) &
\begin{tabular}{@{}c|ccccccc@{}}
$\sigma_y$ [MPa] &
230 & 235 & 245 & 252 & 258 & 262 & 266 \\
$\varepsilon_p$ [--] &
0.0 & 0.0017 & 0.0046 & 0.0064 & 0.0163 & 0.0263 & 0.0362 \\
\end{tabular}
\\
\midrule
\multicolumn{2}{l}{\textbf{Loading program}}\\
Loading type &
Displacement-controlled longitudinal compression \\
Total displacement &
$u = -125\,\mathrm{mm}$ \\
Loading rate &
Constant displacement rate \\
Step duration &
$T = 25\ \mathrm{s}$ \\
\midrule
\multicolumn{2}{l}{\textbf{Numerical parameters}}\\
Solver / step & Abaqus/Explicit; \texttt{*Dynamic, Explicit} \\
Number of nodes / elements & 2082 / 2080 (can); 961 / 900 (top, floor)\\
Element type & 
\begin{tabular}{@{}c|lllllll@{}}
Can &
\texttt{*S4R} shell element, thickness $t=0.05$ [mm], $7$ integration points \\
Top, floor &
\texttt{*R3D4} rigid element \\
\end{tabular}
\\
Contact (core) & Coulomb friction $\mu = 0.3$; hard pressure--overclosure \\
Bulk viscosity & linear $0.06$, quadratic $1.2$ \\
Mass scaling & fixed mass scaling, $\Delta t = 10^{-5}$ \\
\bottomrule
\end{tabularx}
\end{table*}

\begin{thebibliography}{00}
\small
\let\oldbibliography\thebibliography
\renewcommand{\thebibliography}[1]{%
\oldbibliography{#1}%
\setlength{\itemsep}{0pt}%
\setlength{\parsep}{0pt}%
}
\bibitem{ashby1996}
Ashby, M. F., 1996.
Materials Selection in Mechanical Design.
Butterworth-Heinemann.

\bibitem{assran2025vjepa2}
M.~Assran, A.~Bardes, D.~Fan, Q.~Garrido, R.~Howes, M.~Komeili, M.~Muckley, A.~Rizvi, C.~Roberts, K.~Sinha, A.~Zholus, S.~Arnaud, A.~Gejji, A.~Martin, F.~R.~Hogan, D.~Dugas, P.~Bojanowski, V.~Khalidov, P.~Labatut, F.~Massa, M.~Szafraniec, K.~Krishnakumar, Y.~Li, X.~Ma, S.~Chandar, F.~Meier, Y.~LeCun, M.~Rabbat, N.~Ballas,
\newblock V-JEPA 2: Self-Supervised Video Models Enable Understanding, Prediction and Planning,
\newblock arXiv preprint arXiv:2506.09985, 2025.

\bibitem{baillargeon14}
Baillargeon, B., Rebelo, N., Fox, D.D., Taylor, R.L., Kuhl, E., 2014. 
The Living Heart Project: A robust and integrative simulator for human heart function. 
European Journal of Mechanics A/Solids. 48, 38-47.

\bibitem{bendsoe2003}
Bendsøe, M. P., Sigmund, O., 2003.
Topology Optimization: Theory, Methods, and Applications.
Springer.

\bibitem{brown2020}
Brown, T. B., Mann, B., Ryder, N., Subbiah, M., Kaplan, J., Dhariwal, P., Neelakantan, A., Shyam, P., Sastry, G., Askell, A., Agarwal, S., Herbert-Voss, A., Krueger, G., Henighan, T., Child, R., Ramesh, A., Ziegler, D. M., Wu, J., Winter, C., Hesse, C., Chen, M., Sigler, E., Litwin, M., Gray, S., Chess, B., Clark, J., Berner, C., McCandlish, S., Radford, A., Sutskever, I., Amodei, D., 2020.
Language models are few-shot learners.
Advances in Neural Information Processing Systems (NeurIPS), 33, 1877–1901.

\bibitem{coulomb1785}
Coulomb, C. A., 1785.
Théorie des machines simples.
Mémoires de l’Académie Royale des Sciences.


\bibitem{feynman1948}
Feynman, R. P., 1948.
Space-time approach to non-relativistic quantum mechanics.
Reviews of Modern Physics, 20, 367–387.


\bibitem{gibson1979}
Gibson, J. J., 1979.
The Ecological Approach to Visual Perception.
Houghton Mifflin.

\bibitem{goodfellow2014}
Goodfellow, I., Pouget-Abadie, J., Mirza, M., Xu, B., Warde-Farley, D., Ozair, S., Courville, A., Bengio, Y., 2014.
Generative adversarial nets.
Advances in Neural Information Processing Systems (NeurIPS), 27.

\bibitem{grieves2014}
Grieves, M., 2014.
Digital Twin: Manufacturing Excellence through Virtual Factory Replication.
White paper, Florida Institute of Technology.

\bibitem{ha2018world}
Ha, D., Schmidhuber, J., 2018.
Recurrent world models facilitate policy evolution.
Advances in Neural Information Processing Systems (NeurIPS), 31.

\bibitem{hestenes1973}
Hestenes, D., 1973.
The Hamiltonian formulation of classical mechanics.
American Journal of Physics, 41, 905–914.

\bibitem{hirst2026pyvale}
Hirst, J., Sibson, L., Tayeb, A., Poole, B., Sampson, M., Bielajewa, W., Atkinson, M., Marsh, A., Spencer, R., Hamill, R., Hamelin, C., Harte, A., Fletcher, L., 2026.
PYVALE: A Fast, Scalable, Open-Source 2D Digital Image Correlation (DIC) Engine Capable of Handling Gigapixel Images.
arXiv preprint arXiv:2601.12941.

\bibitem{ho2020}
Ho, J., Jain, A., Abbeel, P., 2020.
Denoising diffusion probabilistic models.
Advances in Neural Information Processing Systems (NeurIPS), 33.

\bibitem{ho2022video}
Ho, J., Salimans, T., Gritsenko, A., Chan, W., Dhariwal, P., Chen, M., Sutskever, I., 2022.
Video diffusion models.
arXiv preprint arXiv:2204.03458.

\bibitem{hooke1678}
Hooke, R., 1678.
Lectures de Potentia Restitutiva.
Royal Society, London.

\bibitem{kovachki2023}
Kovachki, N. B., Li, Z., Liu, B., Azizzadenesheli, K., Bhattacharya, K., Stuart, A. M., Anandkumar, A., 2023.
Neural operator: Learning maps between function spaces.
Journal of Machine Learning Research, 24(89), 1–97.

\bibitem{kolmogorov1941}
Kolmogorov, A. N., 1941.
The local structure of turbulence in incompressible viscous fluid.
Doklady Akademii Nauk SSSR, 30, 301–305.

\bibitem{lecun2015}
LeCun, Y., Bengio, Y., Hinton, G., 2015.
Deep learning.
Nature, 521, 436–444.

\bibitem{liu2024physgen}
S.~Liu, Z.~Ren, S.~Gupta, S.~Wang,
\newblock PhysGen: Rigid-Body Physics-Grounded Image-to-Video Generation,
\newblock arXiv preprint arXiv:2409.18964, 2024.

\bibitem{marr1982}
Marr, D., 1982.
Vision.
MIT Press.

\bibitem{mccarthy1955}
McCarthy, J., Minsky, M., Rochester, N., Shannon, C., 1955.
A proposal for the Dartmouth summer research project on artificial intelligence.
Dartmouth College.

\bibitem{moor2023}
Moor, M., Banerjee, O., Abad, Z. S. H., Krumholz, H. M., Leskovec, J., Topol, E. J., Rajpurkar, P., 2023.
Foundation models for generalist medical artificial intelligence.
Nature, 616, 259–265.

\bibitem{moore1965}
Moore, G. E., 1965.
Cramming more components onto integrated circuits.
Electronics, 38, 114–117.

\bibitem{navier1822}
Navier, C. L. M. H., 1822.
Mémoire sur les lois du mouvement des fluides.
Mémoires de l’Académie Royale des Sciences de l’Institut de France.

\bibitem{newton1687}
Newton, I., 1687.
Philosophiæ Naturalis Principia Mathematica.
Royal Society, London.

\bibitem{peirlinck21}
Peirlinck, M., Sahli Costabal, F., Yao, J., Guccione, J.M., Tripathy, S., Wang, Y., Ozturk, D., Segars, P., Morrison, T.M., Levine, S., Kuhl, E, 2021.
Precision medicine in human heart modeling. Perspectives, challenges and opportunities. Biomechanics and Modeling in Mechanobiology, 20, 803-831.

\bibitem{radford2021}
Radford, A., Kim, J. W., Hallacy, C., Ramesh, A., Goh, G., Agarwal, S., Sastry, G., Askell, A., Mishkin, P., Clark, J., Krueger, G., Sutskever, I., 2021.
Learning transferable visual models from natural language supervision.
Proceedings of the 38th International Conference on Machine Learning (ICML), 139, 8748–8763.

\bibitem{saleh15}
Saleh, M., Luzin, V., Toppler, K., Kabir, K., 2015.
Response of thin-skinned sandwich panels to contact loading with flat-ended cylindrical punches: Experiments, numerical simulations and neutron diffraction measurements.
Composites Part B: Engineering, 78, 415–430.

\bibitem{schrodinger1944}
Schrödinger, E., 1944.
What is Life?
Cambridge University Press.

\bibitem{shannon1948}
Shannon, C. E., 1948.
A mathematical theory of communication.
Bell System Technical Journal, 27, 379–423.

\bibitem{sohl2015}
Sohl-Dickstein, J., Weiss, E., Maheswaranathan, N., Ganguli, S., Sohl-Dickstein, J., 2015.
Deep unsupervised learning using nonequilibrium thermodynamics.
Proceedings of the 32nd International Conference on Machine Learning (ICML).

\bibitem{steinmann12}
Steinmann, P., Hossain, M., Possart, G., 2012.
Hyperelastic models for rubber-like materials: consistent tangent operators and suitability for Treloar’s data.
Archive of Applied Mechanics, 82, 1183–1217.

\bibitem{stokes1845}
Stokes, G. G., 1845.
On the theories of the internal friction of fluids.
Transactions of the Cambridge Philosophical Society.

\bibitem{treloar44}
Treloar, L. R. G., 1944.
Stress--strain data for vulcanised rubber under various types of deformation.
Transactions of the Faraday Society, 40, 59–70.

\bibitem{turing1950}
Turing, A. M., 1950.
Computing machinery and intelligence.
Mind, 59, 433–460.

\bibitem{vaswani2017}
Vaswani, A., Shazeer, N., Parmar, N., Uszkoreit, J., Jones, L., Gomez, A. N., Kaiser, {\L}., Polosukhin, I., 2017.
Attention is all you need.
Advances in Neural Information Processing Systems (NeurIPS), 30.

\bibitem{vonneumann1945}
von Neumann, J., 1945.
First draft of a report on the EDVAC.
University of Pennsylvania.

\bibitem{zienkiewicz1977}
Zienkiewicz, O. C., 1977.
The Finite Element Method.
McGraw-Hill.
\end{thebibliography}
\end{document}